\documentclass[times, 10pt,twocolumn]{article}
\usepackage{times}
\usepackage{amssymb}
\usepackage{amsfonts}
\usepackage{amsmath}
\usepackage{bm}
\usepackage{latexsym}
\usepackage{epsfig}

\newtheorem{theorem}{Theorem}[section]

\newtheorem{corollary}[theorem]{Corollary}

\newenvironment{proof}{\noindent \textbf{Proof:}}{$\Box$ \bigskip}

\newcommand{\ignore}[1]{}

\renewcommand{\Pr}{{\bf Pr}}
\renewcommand{\P}{{\bf Pr}}
\newcommand{\Px}{\mathop{\bf Pr\/}}
\newcommand{\Prx}{\mathop{\bf Pr\/}}
\newcommand{\E}{{\bf E}}
\newcommand{\Ex}{\mathop{\bf E\/}}
\newcommand{\Var}{\mathbf{Var}}

\newcommand{\Cov}{\mathbf{Cov}}
\newcommand{\Dev}{\mathbf{Vr}}
\newcommand{\Vr}{\Dev}
\newcommand{\CoDev}{\mathbf{CoVr}}

\newcommand{\Def}{\mathbf{Def}}

\newcommand{\bits}{\{-1,1\}}
\newcommand{\bitsn}{\{-1,1\}^n}
\newcommand{\bitsp}{\{-1,1\}_{(p)}}
\newcommand{\bitspn}{{\bitsp^n}}
\newcommand{\bitsnp}{{\bitsp^n}}
\newcommand{\mup}{\mu_{(p)}}
\newcommand{\hybrid}[3]{#1_{#3}#2}
\newcommand{\fisafunc}{{f: \bitsn \rightarrow \bits}}

\newcommand{\deltap}{\delta}
\newcommand{\Deltap}{\Delta}
\newcommand{\Inf}{\mathbf{Inf}}
\newcommand{\Infmax}{\mathbf{Inf}_{\mathrm{max}}}

\newcommand{\eps}{\epsilon}

\newcommand{\dd}{s}

\newcommand{\Dom}{X}

\newcommand{\CalT}{\mathcal{T}}

\newcommand{\R}{\mathbb R}

\newcommand{\SEL}{\mathrm{SEL}}

\renewcommand{\phi}{\varphi}

\newcommand{\half}{{\textstyle \frac12}}

\begin{document}
\title{Every decision tree has an influential variable}

\author{Ryan O'Donnell\thanks{Some of this research was performed while this author was at the Institute for Advanced Study}\\Microsoft Research\\odonnell@microsoft.com \and Michael Saks\thanks{Supported in part by  NSF grants
CCR-9988526 and CCR-0515201.}\\Rutgers
University\\saks@math.rutgers.edu \and Oded Schramm\\Microsoft
Research\\ \and Rocco A. Servedio\thanks{Supported in part by NSF
CAREER award CCF-0347282 and a Sloan Foundation
Fellowship.
\vskip 0.2in\noindent {\bf
This paper is posted
by permission from the IEEE Computer Society. To appear in FOCS 2005.}
}\\Columbia University\\rocco@cs.columbia.edu
}
\maketitle

\begin{abstract}
We prove that for any decision tree calculating a boolean function
$f:\{-1,1\}^n\to\{-1,1\}$,
\[
\Var[f] \le \sum_{i=1}^n \delta_i\,\Inf_i(f),
\]
where $\delta_i$ is the probability that the $i$th input variable
is read and $\Inf_i(f)$ is the influence of the $i$th variable on
$f$. The variance, influence and probability are taken with
respect to an arbitrary product measure on $\{-1,1\}^n$.
It follows that the minimum depth of a decision tree
calculating a given balanced function is at least the
reciprocal of the largest influence of any input variable.
Likewise, any balanced boolean function with a
decision tree of depth $d$ has a variable with influence at least
$\frac{1}{d}$.  The only previous nontrivial lower bound known was
$\Omega(d 2^{-d})$.  Our inequality has many generalizations,
allowing us to prove influence lower bounds for randomized
decision trees, decision trees on arbitrary product probability
spaces, and decision trees with non-boolean outputs. As an
application of our results we give a very easy proof that the
randomized query complexity of nontrivial monotone graph
properties is at least $\Omega(v^{4/3}/p^{1/3})$, where $v$ is the
number of vertices and $p \leq \half$ is the critical threshold
probability. This supersedes the milestone $\Omega(v^{4/3})$ bound
of Hajnal~\cite{Hajnal:91} and is sometimes
superior to the best known lower bounds of
Chakrabarti-Khot~\cite{ChakrabartiKhot:01} and
Friedgut-Kahn-Wigderson~\cite{FKW:02}.
\end{abstract}

\section{Introduction}
\label{sec:intro}

\subsection{Motivation.}
\label{subsec:mot}
This paper lies at the intersection of two topics within
the theory of boolean functions.

The first topic is {\em decision tree complexity}.  A {\em
deterministic decision tree} (DDT) for a boolean function
$f:\{-1,1\}^n\to \{-1,1\}$ is a deterministic adaptive strategy
for reading variables so as to determine the value of $f$ (a
formal definition appears in Section~\ref{subsec:defs}). The cost
of a DDT on a given input is simply the number of input variables
that it reads, and the DDT complexity of a function $f$, $D(f)$,
is the minimum over all DDT's for $f$ of the maximum cost of any
input. A {\em randomized decision tree} (RDT) for $f$ is a
probability distribution over DDTs for $f$; such trees are
sometimes known as {\em zero-error} randomized decision trees. The
RDT complexity of $f$, $R(f)$, is the minimum over all RDT's for
$f$ of the maximum expected cost of any input. Decision tree
complexity has been studied in theoretical computer science for over 30
years and there is now a significant body of research on the
subject (for a survey, see e.g.,~\cite{BuhrmandeWolf:02}).

The second topic  is {\em variable influences}, introduced to
theoretical computer science by Ben-Or and Linial in
1985~\cite{BenOrLinial:85}. Any $n$-variate boolean function $f$
has an associated {\em influence vector}
$(\Inf_1(f),\ldots,\Inf_n(f))$ where $\Inf_i(f)$ measures the
extent to which the value of $f$ depends on variable $i$ (a
precise definition appears in Section~\ref{subsec:main}). A number
of papers have dealt with properties of this vector and its
relation to other properties of boolean functions;
perhaps the best known work along these lines is that of
Kahn, Kalai and Linial~\cite{KKL:88} (``KKL'') concerning
the maximum influence $\Infmax(f)=\max\{\Inf_i(f):i \in [n]\}$.
Their result implies, for
example, that $\Inf_{\max}(f) = \Omega(\frac{\log n}{n} )$ for any
near-balanced boolean function $f$ (where we say that $f$ is
near-balanced if both $|f^{-1}(1)|/2^n$ and $|f^{-1}(-1)|/2^n$ are
$\Omega(1)$).

The question that originally motivated this paper was: what is the
best lower bound on $\Inf_{\max}(f)$ that holds for all
near-balanced boolean functions $f$ satisfying $D(f) \leq d$?  It
is easy to see that such a function $f$ depends on at most $2^d$
of its variables and therefore the KKL result implies
$\Inf_{\max}(f) \geq \Omega(\frac{d}{2^d})$; prior to this work,
this was the best lower bound known. Our main inequality for
boolean functions, Theorem~\ref{thm:mainboolean}, implies a
(tight) lower bound of $\Inf_{\max}(f) \geq \Omega(\frac{1}{d})$
for any near-balanced function $f$ satisfying $D(f) \leq d$.

In fact, Theorem~\ref{thm:mainboolean} provides a lower bound on a
weighted average of the influence vector, where $\Inf_i(f)$ is
weighted by the probability that a DDT for $f$ queries $x_i$ when
$x$ is a randomly chosen \emph{input}. This lets us extend our
lower bound on $\Infmax(f)$ to functions with $R(f) \leq d$ and
even to functions with $\Delta(f) \leq d$, where $\Delta(f)$
denotes the expected number of queries made by the best DDT for
$f$ on a random input (again, see Section~\ref{subsec:main} for
precise definitions).

\subsection{The main theorem for boolean functions.}
\label{subsec:main}  Our main theorem holds in a very general
setting, that of functions from product probability spaces into
metric spaces.  However the case of greatest interest to us is
much simpler.
Fix some $p\in(0,1)$ and let $\bitspn$ denote the discrete cube endowed with
the $p$-biased product measure, $\mup(x)=p^{|\{i:x_i=1\}|}\,(1-p)^{|\{i:x_i=-1\}|}$.
When we write simply $\bitsn$ the uniform measure case $p = \half$ is implied.
Our main
interest is in boolean functions $f : \bitspn \to \bits$, and in
this section we will describe our main theorem in this case.

First we recall a few definitions.  We have
\[
\Var[f] = \E[f^2] - \E[f]^2 = 4\,\Pr[f = 1]\,\Pr[f = -1].
\]
This measures the ``balance'' of $f$; if $f$ is equally likely to
be $1$ as $-1$, then $\Var[f] = 1$.  We also make the following
definition for the \emph{influence of the $i$th coordinate} on
$f$:
\[
\Inf_i(f) = 2\;\;\Prx_{x, x^{(i)}}[f(x) \neq f(x^{(i)})],
\]
where $x$ is drawn from $\bitsp^n$ and $x^{(i)}$ is formed by
\emph{rerandomizing} the $i$th coordinate of $x$.  Note that
our definition agrees with the one introduced
in~\cite{BenOrLinial:85} in the uniform measure
case $p = \half$, which was $\Inf_i[f] = \Pr[f(x) \neq f(x \oplus
i)]$.  (Our definition differs from the
$p$-biased notion of influences used in,
e.g.,~\cite{FriedgutKalai:96} by a factor of $4\,p\,(1-p)$; we prefer
rerandomizing the $i$th coordinate to flipping it,
since this makes sense in more general product probability spaces which
we will consider later.)  We
call $\Inf(f) := \sum_{i=1}^n \Inf_i(f)$ the \emph{total
influence} of $f$.

Finally, since the notion of influences involves randomizing over
the input domain, it makes sense to introduce a notion of
randomizing over inputs for decision trees.  Let $T$ be a DDT
computing a function $f : \bitsp^n \to \bits$. We write
\vskip -.05in
$$
\deltap_i(T) = \Prx_{x \in \bitsp^n}[\text{$T$ queries $x_i$}],
\quad \quad \quad \text{and}
$$
\vskip -.15in
$$
\Deltap(T) = \sum_{i=1}^n \deltap_i(T) = \Ex_{x \in
\bitspn}[\text{\# coords $T$ queries on $x$ }].
$$
We also let $\Deltap(f)$ denote the minimum of $\Deltap(T)$ over
all DDTs $T$ computing $f : \bitspn \to \bits$.  It is easy to see
that this is equivalent to minimizing over all RDTs computing $f$;
hence $\Deltap(f) \leq R(f)$ for all $p$.  Also note that
$\Deltap(f)$ can be upper-bounded in terms of the \emph{size}
(number of leaves) of the smallest DDT $T$ for
$f$:~\cite{OdonnellServedio:05} shows $\Deltap(f) \leq
\log_2(\text{size(T)})/H(p),$ where $H(p) = -p\log_2
p-(1-p)\log_2(1-p)$ is the binary entropy of $p$.

We may now state our main theorem in the case of functions $f :
\bitsnp \to \bits$:
\begin{theorem} \label{thm:mainboolean}
Let $f : \bitsnp \to \bits$ and let $T$ be a DDT computing $f$.
Then
\vskip -.05in
\[
\Var[f] \leq \sum_{i=1}^n \deltap_i(T)\;\Inf_i(f).
\]
\end{theorem}
As an immediate corollary we
obtain the lower bound on $\Infmax(f)$ mentioned in
Section~\ref{subsec:mot}:
\begin{corollary}
\label{cor:Imaxboolean}
For every
$f : \bitsnp \to \bits$ we have
\[\Deltap(f)\geq \frac{\Var(f)}{\Infmax(f)}\,.\]
\end{corollary}
\noindent {\bf Proof:}  Let $T$ be a DDT computing $f$. From
Theorem~\ref{thm:mainboolean},
\begin{multline*}
\Var[f] \leq \sum_{i=1}^n \deltap_i(T)\;\Inf_i(f) \\ \leq
\Infmax(f) \sum_{i=1}^n \deltap_i(T) = \Infmax(f)\cdot\Deltap(T)\,
. \quad \Box
\end{multline*}

Some brief comments on our main theorem:
 \begin{itemize}
  \item It is linear in the $\deltap_i(T)$'s.  Hence if we allow
  an RDT $\CalT$ for $f$ and make the natural definition of
  $\deltap_i(\CalT)$, the result still holds by averaging over the
  distribution $\CalT$.
  \item It can be sharp; see Section~\ref{subsec:tightness} for cases of equality.
  \item Other corollaries along the lines of
  Corollary~\ref{cor:Imaxboolean} follow; for example, if $d$ is
  an integer $\ge \Delta(f)$, then the sum of the influences of the $d$ most
  influential variables is at least $\Var[f]$.
  \item In Section~\ref{subsec:twofunc} we will give a ``two function'' version, which yields
  a lower bound for the
  randomized decision tree complexity of approximating $f$.
 \end{itemize}

\subsubsection{Influence lower bounds --- comparison with previous work.}
\label{sec:comparison} Proving lower bounds on the influences of
boolean functions has had a long history in theoretical computer
science, starting with the 1985 paper of Ben-Or and
Linial~\cite{BenOrLinial:85} on collective coin flipping.
Ben-Or
and Linial made the basic observation that if $f : \bitsn \to
\bits$ is balanced (i.e., $\E[f]=0$), then $\Infmax(f) \geq \frac{1}{n}$.  This
follows from the edge isoperimetric inequality on the discrete
cube (see, e.g.,~\cite{Bollobas:86}); however, it is more
instructive for us to view it as following from the
\emph{Efron-Stein inequality} \cite{EfronStein:81,Steele:86},
\begin{equation} \label{eqn:efronstein}
\Var[f] \leq \Inf(f) = \sum_{i=1}^n \Inf_i(f),
\end{equation}
which holds in the general $p$-biased case, and also in the much
more general setting of $f : \Omega \to \R$, where $\Omega$ is a
$n$-wise product probability space and $\Inf_i$ is defined
appropriately for real-valued functions (specifically, with the
``$\rho_2$ semimetric'' discussed in
Section~\ref{subsec:notametric}).  Theorem~\ref{thm:mainboolean}
is immediately seen to improve the Efron-Stein inequality in the
case of functions $f : \bitspn \to \bits$.

Ben-Or and Linial constructed a balanced function $f : \bitsn \to
\bits$ (``Tribes'') satisfying $\Infmax(f) = \Theta(\frac{\log
n}{n})$ and conjectured that for every balanced function
$f:\bitsn\to\bits$, $\Infmax$ cannot be smaller.
   There were
small improvements on the simple $\frac{1}{n}$ bound
($\frac{2-\eps}{n}$ by Alon, $\frac{3-\eps}{n}$ by Chor and
Gereb-Graus; see~\cite{KKL:88}) before the famous KKL
paper~\cite{KKL:88} confirmed the conjecture.  Note that our
theorem improves upon KKL whenever $f$ has $\Delta(f) = o(n/\log
n)$; in particular, whenever $f$ has a DDT of size $2^{o(n/\log
n)}$.

The KKL result was subsequently generalized by
Talagrand~\cite[Theorem 1.5]{Talagrand:94} who proved that for any $f : \bitsnp
\to \bits$,
\begin{equation} \label{eqn:tal}
\Var[f] \leq O\Bigl(\log\frac{1}{p(1-p)}\Bigr) \sum_{i=1}^n \frac{
\Inf_i(f) }{\log(1/\Inf_i(f))}\,.
\end{equation}
Talagrand's motivation for proving this was that when $f : \bitspn
\to \bits$ is monotone, lower bounds on the sum of $f$'s
influences imply a ``sharp threshold'' for $f$, via the
Russo-Margulis lemma~\cite{Margulis:74,Russo:78}.  Indeed, this
connection with threshold phenomena is one of the chief motivations
for studying influences, and it is considered an important
problem in the theory of boolean functions and random graphs to
provide general conditions under which the total influence is
large~\cite{BourgainKalai:97}. Our main inequality provides such a
condition: $\Inf(f)$ is large if $f$ has a randomized decision
tree $\CalT$ with $\deltap_i(\CalT)$ small for all $i$. Note that
when $f$ is a transitive function, this is equivalent to the
natural condition that $\Deltap(f)$ is small. (See
Section~\ref{sec:RDT} for definitions of monotone and transitive
functions, as well as further discussion of random graph
properties.)

In particular, ours seems to be the first quantitatively strong
influence lower
bound that takes into account the ``structure'' or computational
complexity of $f$.  We note that previously achievable lower
bounds on influences in terms of some measure of the complexity of
$f$ yield quantitatively much weaker results than can be obtained
from our inequality.  For instance, Nisan and Szegedy
\cite{NisanSzegedy:92} showed that if $f : \bitsn \to \bits$ is
computed by a polynomial over $\R$ of degree deg$(f)$, then every
coordinate~$i$ with nonzero influence has $\Inf_i(f) \geq
2^{-\text{deg}(f)}$.  Since $D(f) \leq O($deg$(f)^4)$ (by a result
of Nisan and Smolensky~\cite{BuhrmandeWolf:02}), our
Corollary~\ref{cor:Imaxboolean} implies that the maximum influence
in fact satisfies $\Infmax(f) \geq \Omega(
\Var[f]/\text{deg}(f)^4)$. As another example, suppose $\fisafunc$ is
approximately computed by a polynomial over $\R$ of degree
$\widetilde{\text{deg}}(f)$ --- i.e.\ there is a polynomial $p(x)$
of degree $\widetilde{\text{deg}}(f)$ such that $|p(x) - f(x)| <
1/3$ for all $x$.  Talagrand's result implies that $\Infmax(f)
\geq \exp(-O(\Inf(f)/\Var[f]))$.  Since by~\cite{Shi:00} we have
$\Inf(f) \leq O(\widetilde{\text{deg}}(f))$, one could conclude
that $\Infmax(f) \geq
\exp(-O(\widetilde{\text{deg}}(f)/\Var[f]))$. However by contrast,
since $D(f) \leq O(\widetilde{\text{deg}}(f)^6)$
by~\cite{BBC+:01}, our Corollary~\ref{cor:Imaxboolean} implies
that the maximum influence in fact satisfies $\Infmax(f) \geq
\Omega(\Var[f]/\widetilde{\text{deg}}(f)^6)$.

\section{Randomized decision tree complexity lower bounds}
\label{sec:RDT}

In this section we give an
application of Theorem~\ref{thm:mainboolean} to the problem of
randomized decision tree complexity for monotone graph properties.
We prove Theorem~\ref{thm:mainboolean} in a more general setting
in Section~\ref{sec:main}.

\subsection{History.} \label{subsec:RDT-history}
As mentioned in Section~\ref{subsec:mot}, decision tree complexity
has been extensively studied for over three decades.  Two special
classes of functions have played a prominent role in these
investigations. The first is the class of {\em monotone
functions}, those satisfying $f(y) \geq f(x)$ whenever $y \geq x$
under the componentwise partial order. The second is the class of
{\em transitive functions}.  An automorphism of the $n$-variate
boolean function $f$ is a permutation $\sigma$ of $[n]$ satisfying
$f(x_1,\dots,x_n)=f(x_{\sigma(1)},\dots,x_{\sigma(n)})$ for all
inputs $x$.  We say that $f$ is {\em transitive} if for each pair
$i,j \in [n]$ there is an automorphism of $f$ that sends  $i$ to
$j$.  For example, Rivest and Vuillemin~\cite{RivestVuillemin:76}
proved that for $n$ a prime power, any $n$-variate monotone
transitive function $f$ has $D(f)=n$.

One long studied open question about boolean decision tree
complexity is the following: how small can $R(f)$ be in relation
to $D(f)$? It is well known~\cite{BlumImpagliazzo:87} that $R(f)
\geq \Omega(\sqrt{D(f)})$ for any function $f$, and this is the
best general lower bound known. The largest known separation is
given by the following recursively defined function: Let $f_0$ be
the identity function on a single variable and for $k \geq 1$, let
$f_k$ be the function on $n=4^k$ variables given by $(f_{k-1}^1
\wedge f_{k-1}^2) \vee (f_{k-1}^3 \wedge f_{k-1}^4)$, where
$f_{k-1}^i$ is the value of $f_{k-1}$ on the $i$th group of
$4^{k-1}$ variables. The function $f_k$ is monotone and
transitive, and so by the above result of Rivest and Vuillemin,
$D(f_k)=n$.  Snir~\cite{Snir:85} gave an RDT for $f_k$
establishing $R(f) \leq n^{\beta}$ where $\beta=\log_2 \left(
{\frac {1 + \sqrt{33}}4}\right) \approx 0.753$. Saks and Wigderson
\cite{SaksWigderson:86} proved that Snir's RDT is optimal for
$f_k$ and conjectured that $R(f) \geq \Omega(D(f)^{\beta})$ for
any boolean function; this is not even known to hold for all
monotone transitive functions.

A well studied subclass of transitive boolean functions consists
of functions derived from graph properties.  A {\em property} of
$v$-vertex (undirected) graphs is a set of graphs on vertex set
$V=\{1,\dots,v\}$ that is invariant under vertex relabellings;
e.g., the set of graphs on $V$ that are properly 3-colorable.  We
restrict attention to properties that are non-trivial; i.e., at
least one graph has the property and at least one graph does not
have the property.

Let $\binom{V}{2}$ denote the set of 2-elements subsets of $V$.
Each graph $G$ on $V$ can be identified with the boolean vector
$x^G\in \bits^{\binom{V}{2}}$ where $x^G_{\{i,j\}}$ is $1$ if
$\{i,j\} \in E(G)$ and is $-1$ otherwise. A graph property
$\mathcal{P}$ is thus naturally identified with a boolean function
$f_{\mathcal{P}}:\bits^{\binom{V}{2}} \longrightarrow \bits$ which
maps the vector $x^G$ to 1 if and only if $G$  satisfies
$\mathcal{P}$.  The invariance of  properties under vertex
relabellings implies that the associated functions are transitive.

There are examples of graph properties on $v$ vertices that have
deterministic decision trees of depth $O(v)$; e.g., the property
of being a ``scorpion graph''~\cite{BVL:74}.  However, for graph
properties that are {\em monotone} (those whose associated
function is monotone), Rivest and
Vuillemin~\cite{RivestVuillemin:76} proved a lower bound
$\Omega(v^2)$ on DDT complexity.  A conjecture made by Yao
\cite{Yao:77} and also attributed to Karp \cite{SaksWigderson:86}
is that this $\Omega(v^2)$ lower bound
extends to RDT complexity.  This is the problem we make
progress on in this section.

Yao observed that an $\Omega(v)$ lower bound for RDT computation
of monotone graph properties is easy to prove; this also
follows from the general bound $R(f)=\Omega(\sqrt{D(f)})$
mentioned earlier.  The first improvement on this naive bound came
a decade later from Yao himself, who proved an $\Omega(v
\log^{1/12} v)$ lower bound using ``graph packing''
arguments~\cite{Yao:87}. These arguments were improved by
King~\cite{King:88}, yielding an $\Omega(v^{5/4})$ lower bound, and
by Hajnal~\cite{Hajnal:91}, yielding an $\Omega(v^{4/3})$ lower
bound. This lower bound stood for a decade before Chakrabarti and
Khot~\cite{ChakrabartiKhot:01} gave a small improvement to
$\Omega(v^{4/3} \log^{1/3} v)$.  Both the Hajnal and
Chakrabarti-Khot bounds have rather long and technical proofs
based on graph packing.

Fairly recently, Friedgut, Kahn and Wigderson~\cite{FKW:02} proved
a general lower bound of a somewhat different form.  Given a
nonconstant monotone boolean function $f : \bitsnp \to \bits$, it
is easy to see that $\E[f]$ is a continuous increasing function of
$p$; therefore there is a \emph{critical probability} $p$ for
which $\E[f] = 0$, i.e., $\Var[f] = 1$.  Friedgut, Kahn and
Wigderson proved that any nontrivial monotone $v$-vertex graph
property has RDT complexity $\Omega
(\min\{\frac{v}{\min(p,1-p)},\frac{v^2}{\log v}\})$ when $p$ is
the critical probability for $f$.  In fact, they show that
$\Deltap(f)$ is at least this quantity. The FKW bound can improve
on Chakrabarti-Khot in cases where the critical probability is
sufficiently close to 0 or 1.  We remark that the proof in FKW
also uses a graph packing argument.

\subsection{Our $R(f)$ lower bound.} \label{subsec:our-RDT}
As a simple consequence of our elementary main inequality
Theorem~\ref{thm:mainboolean} and a recent elementary inequality
from~\cite{OdonnellServedio:05}, we obtain the following:
\begin{theorem}
\label{thm:rdt lb} Let $f : \bitsnp \to \bits$ be a nonconstant
monotone transitive function, where $p$ is the critical
probability for $f$ (i.e., $f$ is balanced).  Write $q = 1-p$.  Then
 \[
  R(f) \geq \Deltap(f) \geq \frac{n^{2/3}}{(4\,p\,q)^{1/3}}.
 \]
In particular,
\[
R(f) \geq  \Deltap(f) \geq \frac{(v-1)^{4/3}}{(16\,p\,q)^{1/3}}
 \]
if $f$ corresponds to a $v$-vertex graph property.
\end{theorem}
\begin{proof}
The inequality we need from~\cite{OdonnellServedio:05} is the
following:\\

For all $p$, if $f : \bitsnp \to \bits$ is monotone then
\begin{equation} \label{eqn:OS}
\Inf(f) \leq 2\sqrt{p\,q\,\Deltap(f)}\,.
\end{equation}
Fix $p$ to be the critical probability of $f$ and let $T$ be a
DDT computing $f$ with expected cost $\Deltap(f)$.  We apply
Theorem~\ref{thm:mainboolean}, using $\Var[f] = 1$ since $p$ is
critical and $\Inf_i(f) = \Inf(f)/n$ since $f$ is transitive (and
hence all coordinates have the same influence).  This gives $1
\leq (\Inf(f)/n)\cdot\Deltap(f)$. Using~(\ref{eqn:OS}) to bound
$\Inf(f)$ we get $1 \leq (2\sqrt{p\,q}/n)\cdot(\Deltap(f))^{3/2}$,
and this can be rearranged to give the desired result.
\end{proof}

\subsection{Discussion.}  In the case of monotone graph
properties, our result always improves on Hajnal's
$\Omega(v^{4/3})$ lower bound and can be superior to both
Chakrabarti-Khot (when $\min\{p,q\}$ is small enough) and to FKW
(when $\min\{p,q\}$ is large enough).  It is worth noting
that unlike all previous lower bounds for monotone
graph properties, our proof makes no use of
graph packing arguments, instead relying only on
elementary probabilistic arguments.

Most interestingly, we obtain a result essentially as
good as the best unconditional bound (Chakrabarti-Khot) in the
more general context of monotone \emph{transitive} functions, not
just graph properties.  Further, our bound for
monotone transitive functions is
known to be essentially tight in the case $p = 1/2$:
in~\cite{BSW:05}, a sequence $f_n:\bitsn\to\bits$ of balanced monotone
transitive functions is presented with $\Delta(f_n) \leq O(n^{2/3}
\log n)$.
Our present Theorem~\ref{thm:mainboolean} is used in~\cite{BSW:05}
to show that $\Delta(f_n)=\Omega(n^{2/3})$, by an argument similar
to the proof of
Theorem~\ref{thm:rdt lb}, but using an inequality
from~\cite{SchrammSteif:05} in place of~(\ref{eqn:OS}).

It is tantalizing that the place where the RDT complexity
of monotone graph properties has been stuck for almost 15 years,
$v^{4/3}$, is exactly the tight bound for monotone transitive
functions. Perhaps this suggests that in some way the argument of
Hajnal is not really using the fact that $f$ is a graph property
--- just that it's transitive.  Indeed, one might wonder the same thing
about Chakrabarti-Khot, since their $v^{4/3} \log^{1/3} v$ lower
bound could also hold for monotone transitive functions --- the
example of~\cite{BSW:05} does not rule it out.

\section{The main inequality}
\label{sec:main}
\subsection{Decision trees, variation, influences --- general definitions.}
\label{subsec:defs}

The proof of Theorem~\ref{thm:mainboolean} is
most naturally carried out in a significantly more general context
than that of functions $f : \bitspn \to \bits$.  Specifically, we
will consider functions
\[
f : \Omega \longrightarrow Z
\]
mapping a \emph{product probability space} into a \emph{metric
space}.  In this section we give the necessary definitions.

Let us begin with the domain.  Here we have an $n$-wise product
probability space $\Omega=(\Dom,\mu)$, meaning that that the
underlying set $\Dom$ is a product set $\Dom_1 \times \cdots
\times \Dom_n$ and the measure $\mu$ is a product probability
measure $\mu_1 \times \cdots \times \mu_n$, where $\mu_i$ is a
probability measure on $\Dom_i$.
For simplicity, we assume that $X$ is finite.
We write $\Omega_i$ for the probability space $(\Dom_i,\mu_i)$. We
use the notation $x \leftarrow \Omega$ to mean that $x$ is an
element of $\Dom$ randomly selected according to $\mu$.

The range of our functions is a metric space $(Z,d)$.
(Actually we can allow a ``pseudo-metric'', meaning we may omit the requirement
that $d(z,z') = 0 \Rightarrow z = z'$.)  Useful examples to
keep in mind are the following: $Z$ any finite set with $d(z,z') =
{\bf 1}_{z \ne z'}$; and, $Z = \R$ with $d(z,z') = |z - z'|$.  Of
course, in the special case of boolean-valued functions, $Z =
\bits$, all metrics are the same up to a constant factor.

The definitions of decision trees in the context of functions
mapping a product set domain $\Dom = \Dom_1 \times \cdots \times
\Dom_n$ into a set $Z$ are the obvious ones.  Briefly, a DDT will
be a rooted directed tree $T$ in which each internal node $v$ is
labelled by a coordinate $i_v \in [n]$ and each leaf is labelled
by an element of the output set $Z$. Further, the arcs emanating
from each internal node $v$ must be in one-to-one correspondence with
$\Dom_{i_v}$. The node labels along every root-leaf path are
required to be distinct.  $T$ computes a function $f_T : \Dom \to
Z$ in the obvious way; we retain the notion of the cost of $T$ on
input $x$ as the length of the root-leaf path $T$ follows on input
$x$.  Thus, we have the usual notions of $D(T)$ and $D(f)$, and
also the (zero-error) randomized decision tree complexities
$R(\CalT)$ and $R(f)$.  With the product probability measure $\mu$
on $\Dom$, we can also naturally extend our notions of
\emph{expected cost} from Section~\ref{subsec:main}: given a DDT
$T$ computing $f$,
\[
\delta_i^{\mu}(T) = \Px_{x \leftarrow \Omega %
}[\text{$T$ queries $x_i$}],
\]
and $\Delta^{\mu}(T)$ and $\Delta^{\mu}(f)$ are similarly defined.
We will henceforth drop the superscript $\mu$ when it is clear
from context.  Note that, as before, we have $\Delta(f) \leq R(f)$.

We now give the definitions of \emph{variation} and
\emph{influences} for functions $f : \Omega \to Z$.  The
\emph{variation} of $f : \Omega \to Z$ is
\[
\Dev^{\mu, d}[f] = \Ex_{(x,y) \leftarrow \Omega \times
\Omega}[d(f(x),f(y))].
\]
To define influences, first let $\Omega^{(i)}$ denote the
probability space given by pairs $(x,x^{(i)})$, where $x$ is
chosen from $\Omega$ and $x^{(i)}$ is formed by rerandomizing the
$i$th coordinate of $x$ using $\mu_i$.  Then the \emph{influence
of the $i$th coordinate} on $f : \Omega \to Z$ is defined to be
\[
\Inf^{\mu,d}_i(f) = \Ex_{(x,x^{(i)}) \leftarrow
\Omega^{(i)}}[d(f(x),f(x^{(i)}))].
\]
We will usually drop the superscripts $\mu$ and $d$ on $\Dev$ and
$\Inf_i$ when they are implied by context.  Note that if we view
the functions $f : \bitsnp \to \bits$ from Section~\ref{sec:intro}
as mapping into the metric space on $\bits$ with distance $d$
given by $d(z,z') = |z - z'| = 2\cdot{\bf 1}_{z \neq z'}$, then we
get agreement in the definitions of $\Inf_i(f)$ and also $\Dev[f]
= \Var[f]$.

\subsection{Theorem and proof.} \label{subsec:proof}
We now state and prove our main inequality, which includes
Theorem~\ref{thm:mainboolean} as a special case.
\begin{theorem} \label{thm:main}
Let $f : \Omega \to (Z,d)$ be a function mapping a finite $n$-wise
product probability space into a metric space, and let $T$ be a
DDT computing $f$.  Then
\[
\Dev[f] \leq \sum_{i=1}^n \delta_i(T)\;\Inf_i(f).
\]
\end{theorem}
\begin{proof}
Let $x$ and $y$ be random inputs chosen independently from
$\Omega$.  Given a subset $J \subseteq [n]$ we will write
$\hybrid{x}{y}{J}$ for the \emph{hybrid input} in $\Dom$ that
agrees with $x$ on the coordinates in $J$ and with $y$ on the
coordinates in $[n] \setminus J$.  Let $i_1, \dots, i_\dd$ denote
the sequence of variables queried by $T$ on input $x$ (these $i$'s
are random variables and $\dd$ is also a random variable).  For $t
\geq 0$, let $J[t] = \{i_r : \dd\ge r > t\}$.  Finally, let $u[t] =
\hybrid{x}{y}{J[t]}$.
(For example, if $x=(1,-1,1,1)$, $y=(1,1,-1,-1)$,
the tree read $x_4$ followed by $x_2$ and terminates, then
$u[0]=(1,-1,-1,1)$, $u[1]=(1,-1,-1,-1)$ and $u[2]=y$.)
  All $\E[\,\cdot\,]$'s and $\Pr[\,\cdot\,]$'s in
what follows are over all the random variables just described
(i.e., $x$, $y$, $i$'s, $\dd$, $u[\cdot]$'s).

We begin with the simple observation
\[
\Dev[f] = \E[d(f(x),f(y))] = \E[d(f(u[0]),f(u[\dd]))],
\]
which follows because $y = u[\dd]$ and $f(x) = f(u[0])$ (although
$x$ does not necessarily equal $u[0]$).  This latter equality is
the only place in the proof we use the fact that $T$ computes $f$.

We next make the obvious step
\begin{equation}\label{ieq}
\E[d(f(u[0]),f(u[\dd]))] \leq \E\Bigl[\sum_{t = 1}^s
d(f(u[t-1]),f(u[t]))\Bigr]
\end{equation}
which uses the fact that $d$ is a metric.
Set $i_t=\emptyset$ for $t>s$.
Linearity of expectation and
$1_{\{t\le s\}}=\sum_{i=1}^n 1_{\{i_t=i\}}$ give
\begin{multline}
\E\Bigl[\sum_{t =1}^s d(f(u[t-1]),f(u[t]))\Bigr] = \\
\sum_{t = 1}^n\sum_{i = 1}^n \E\Bigl[d\bigl(f(u[t-1]),f(u[t])\bigr)\,1_{\{i_t=i\}}\Bigr].
\label{eq:condition}
\end{multline}

Let $X_t$ denote
the sequence of values seen by the decision tree
by time $t$ on input $x$; that is,
$X_t=(x_{i_1}, \dots, x_{i_{t\wedge s}})$,
where $t\wedge s$ denotes the minimum of $t$ and $s$.
Note that $X_{t-1}$ determines $i_t$.
Induction on $t\in[n]$ easily shows that conditional on $X_{t-1}$
the variables $y$ and $(x_j:j\ne i_1,\dots,i_{(t-1)\wedge s})$
are independent and retain their original distributions.
It follows that
conditional on $X_{t-1}$
the pair $(u[t-1],u[t])$ has the distribution $\Omega^{(i)}$
if $i_t=i\in[n]$.
Consequently, for $i,t\in[n]$,
$$
\E\Bigl[d\bigl(f(u[t-1]),f(u[t])\bigr)\,1_{\{i_t=i\}} \Bigm|X_{t-1}\Bigr]
=
1_{\{i_t=i\}} \,\Inf_i(f)\,.
$$
Taking expectation gives
$$
\E\Bigl[d\bigl(f(u[t-1]),f(u[t])\bigr)\,1_{\{i_t=i\}} \Bigr]
=\P[i_t=i]\,\Inf_i(f)\,.
$$
Since $\sum_{t=1}^n\P[i_t=i]=\delta_i(T)$, an appeal to~(\ref{eq:condition})
completes the proof.
\end{proof}

\subsection{Corollaries and two function version.}
\label{subsec:twofunc} In this section we treat some immediate
corollaries of Theorem~\ref{thm:main}.  Certainly the analogue of
Corollary~\ref{cor:Imaxboolean} holds for Theorem~\ref{thm:main},
as do the first and third remarks stated after
Corollary~\ref{cor:Imaxboolean}.
  We now give the promised ``two
function'' version.  Define
 \begin{multline*}
\CoDev[f,g] \\
= \Ex_{(x,y) \leftarrow \Omega \times \Omega}[d(f(x),g(y))] -
\Ex_{x \leftarrow \Omega}[d(f(x),g(x))],
\end{multline*}
so in particular $\CoDev[f,f] = \Dev[f]$.  Thus the following
theorem generalizes Theorem~\ref{thm:main}:

\begin{theorem} \label{thm:twofunc}
Let $f, g : \Omega \to (Z,d)$ be functions mapping a finite $n$-wise
product probability space into a metric space, and let $\CalT$ be
an RDT computing $f$.  Then
\[
\bigl|\CoDev[f,g]\bigr| \leq \sum_{i=1}^n
\delta_i(\CalT)\;\Inf_i(g).
\]
\end{theorem}
\begin{proof}
As usual we can assume by averaging that $\CalT$ is a DDT $T$
computing $f$.  Using the same setup as in the proof of
Theorem~\ref{thm:main}, we have
\begin{multline*}
\CoDev[f,g] = \E[d(f(x),g(y))] - \E[d(f(u[0]),g(u[0]))] \\=
\E[d(f(u[0]),g(u[s]))] - \E[d(f(u[0]),g(u[0]))]
\end{multline*}
where in the first equality we used that $u[0]$ is, in isolation,
distributed according to $\Omega$, and in the second equality we
used the fact that $f(x) = f(u[0])$ since $T$ computes $f$ (as in
the previous proof). Now using the fact that $d$ is a metric we
get
\begin{multline*}
\CoDev[f,g] = \\ \E[d(f(u[0]),g(u[s]))] - \E[d(f(u[0]),g(u[0]))]
\\
\leq \E[d(g(u[0]),g(u[s]))]
\end{multline*}
and of course this is also true for $-\CoDev[f,g]$.  The proof now
proceeds exactly as before with $g$ in place of $f$; note that
from this point on in the previous proof we did not use the fact
that $T$ computed $f$.
\end{proof}

As mentioned below Corollary~\ref{cor:Imaxboolean},
Theorem~\ref{thm:twofunc} can be used to give a lower bound for the
randomized decision tree complexity of approximating $g$.
Note that the triangle inequality gives
$$
\CoDev[f,g] \ge \Vr[g]-2\,\E\bigl[d(f(x),g(x))\bigr].
$$
Consequently, Theorem~\ref{thm:twofunc} implies
that for every $\epsilon>0$ the expected number of queries required
by a randomized decision tree to calculate
any approximation $f$ of $g$
satisfying $\E\bigl[d(f(x),g(x))\bigr]\le\epsilon$
is at least
$$
\frac{\Vr[g]-2\,\epsilon}{\Infmax(g)}\,.
$$
\bigskip

We now describe an alternate version of Theorem~\ref{thm:twofunc}.
Let $f:\Omega \to [-1,1]$, $g:\Omega\to \R$,
 and let $\CalT$ be a randomized decision tree computing $f$.  Then
\begin{equation}\label{e:alt}
\bigl|\Cov[f,g]\bigr| \le \sum_{i=1}^n \delta_i(\CalT)\,
\Inf_i^{\rho_1}[g],
\end{equation}
where $\rho_1(x,y)=|x-y|$ and $\Cov[f,g]=\E[f(x)\,g(x)]-\E[f(x)]\,\E[g(x)]$
is the covariance of $f$ and $g$.
With the usual definitions of $x,y,u[t]$ and $\dd$, we have
$f(x)=f[u(0)]$ and $u[\dd]=y$. Hence, we may write
\begin{multline*}
\Cov[f,g] = \E\bigl[f(u[0])g(u[0])-f(x)g(u[\dd])\bigr] \\ =
\E\bigl[f(x)\bigl(g(u[0])-g(u[\dd])\bigr)\bigr] \le
 \E\bigl[|g(u[0])-g(u[\dd])|\bigr]\,,
\end{multline*}
and the proof of~(\ref{e:alt}) proceeds as above.

\subsection{When $d$ is not a metric.} \label{subsec:notametric}
In this section we generalize our results to the case when $f$
maps into $(Z, \rho)$, where $(Z,\rho)$ is a ``semimetric''.  This
just means that $\rho$ need not satisfy the triangle inequality;
specifically, all we require of $\rho$ is that $\rho \geq 0$,
$\rho(z,z)=0$, and $\rho(z,z') = \rho(z',z)$.  (Again we do not
insist that $\rho(z,z') = 0 \Rightarrow z = z'$.)  Our main
motivation for studying this extension is the case $Z = \R$ with
$\rho = \rho_2(z,z') := (z-z')^2/2$.  In this case
$\Dev^{\rho_2}[f] = \Var[f]$ and $\Inf^{\rho_2}(f)$ has the
meaning commonly associated with this notation for functions $f :
\Omega \to \R$; that is, the interpretation used in, e.g., the
Efron-Stein inequality or in~\cite{MOO:05}. 

To study the semimetric case, we simply introduce a quantity
measuring the extent to which the triangle inequality fails for
$\rho$ on paths of length $k$.  We define the {\em defect} of a
sequence $z_0,z_1,\dots,z_k \in Z^{k+1}$ to be
$\rho(z_0,z_k)\bigm/(\sum_{t=1}^k \rho(z_{t-1},z_t))$, where
$\frac{0}{0}$ is taken to be 1. We then define the {\em $k$-defect
of $\rho$}, denoted $\Def_k(\rho)$, to be the maximum defect of
any sequence $z_0,\ldots,z_{k}$.  The following facts are easy
to check:
\begin{itemize}
 \item $\Def_1(\rho)=1$ and $\Def_k(\rho)$ is nondecreasing with $k$.
 \item $\Def_k(\rho) \leq (\sup \rho)/(\inf \rho)$ for all $k$.
 \item $\Def_2(\rho)=1$ implies that $\rho$ satisfies the triangle inequality,
       which, in turn, implies that $\Def_k(\rho)=1$ for all $k$;
       i.e., $\rho$ is a metric.
 \item If $\rho^{1/q}$ is a metric for some $q \geq 1$, then $\Def_k(\rho) \leq
       k^{q-1}$. Thus in our motivating case with $Z = \R$ and $\rho(z,z') =
       (z-z')^2/2$ we have $\Def_k(\rho) \leq k$.
 \item If $\rho^{1/q}$ is a metric for some $q \geq 1$,
       then $\Def_k(\rho) \leq |Z|^{q-1}$ for all $k$.
\end{itemize}
It is easy to see how to generalize Theorems~\ref{thm:main}
and~\ref{thm:twofunc} for semimetrics $\rho$; since
Theorem~\ref{thm:twofunc} is more general, we will only state its
extension:
\begin{theorem} \label{thm:semimetric}
Let $f, g : \Omega \to (Z,\rho)$ be functions mapping an $n$-wise
product probability space into a semimetric space, and let $\CalT$
be an RDT computing $f$.  Let $k$ be the length of the longest
path in any DDT in $\CalT$'s support.  Then
\[
\bigl|\CoDev[f,g]\bigr| \leq \Def_k(\rho)\;\sum_{i=1}^n
\delta_i(\CalT)\;\Inf_i(g).
\]
\end{theorem}
This is the most general version of our main inequality that we
state. In the semimetric setting we are most interested in, namely
that of one function $f : \Omega \to (\R, \rho_2)$, we have the
following:
\begin{corollary} \label{cor:real}
Let $f : \Omega \to (\R, \rho_2)$ be a function mapping an
$n$-wise product probability space into the real line with
semimetric $\rho_2(z,z') = (z-z')^2/2$, and let $\CalT$ be an RDT
computing $f$. Let $k$ be the length of the longest path in any
DDT in $\CalT$'s support. Then
\[
\Var[f] \leq k\;\sum_{i=1}^n \delta_i(\CalT)\;\Inf^{\rho_2}_i(f),
\]
and $f$ has a coordinate with $\rho_2$-influence at least $\Var[f]/k^2$
(since $\sum_{i=1}^n \delta_i(\CalT)\le k$).
\end{corollary}

\subsection{Tightness of the inequality} \label{subsec:tightness}
Our main Theorem~\ref{thm:main} can be tight; one class of
DDTs for which it is tight are \emph{read-once} decision trees.
In fact, it is tight for a broader family of decision trees, which
we now describe. Observe that each subtree of a decision tree
below any given node can be thought of as a decision tree on the
same input $\Omega$ (which may ignore some of the input variables).
Say that a decision tree is \emph{separated}, if for every
two subtrees $T'$ and $T''$ and every input $x\in \Omega$,
if $T'$ and $T''$ compute different values on $x$, then
the sets of variables they query on input $x$ are disjoint.
Clearly, read-once trees are separated. Later, we will see
that separated trees are not necessarily read-once.

To prove that Theorem~\ref{thm:main} is tight for every
separated tree, note that
 the only inequality in the proof of the theorem is (\ref{ieq}).
Suppose that $T$ is separated, that
$f(u[0])\ne f(u[t])$ and that $t$ is minimal with this property.
Since $f(u[t])\ne f(u[t-1])$, on input $u[t]$ the variable
$y_{i_t}$ is inspected by $T$.
Let $v'$ be the node of $T$ arrived at
right after reading $x_{i_{t}}$ on input $x$, and let
$v''$ be the node of $T$ arrived at right after reading
$y_{i_{t}}$ on input $u[t]$. Then on input $u[t]$
the two subtrees of $T$ rooted at $v'$ and $v''$
calculate $f(x)=f(u[0])$ and $f(u[t])$, respectively.
Since $f(u[t])\ne f(u[0])$ and $T$ is separated,
the sets of variables examined by these two subtrees on
input $u[t]$ are disjoint. In particular,
$f(u[t])=f(u[t+1])=\cdots=f(u[s])$.
Since $f(u[0])=f(u[1])=\cdots=f(u[t-1])$,
this shows that~(\ref{ieq}) must hold as an equality
when $T$ is separated. Thus, the inequality in Theorem~\ref{thm:main}
holds as an equality in this case.
Note that this argument shows
that equality holds even if $d =
\rho$ is just a semimetric.

Simple examples of read-once DDTs are those for AND $:\bits^n \to
\bits$ and OR $:\bits^n \to \bits$.
The simplest nontrivial balanced example is the
``selection function'' $\SEL:\bits^3 \to \bits$, which maps
$(x_1, x_2, x_3)$ to $x_2$ if $x_1 = 1$, or $x_3$ if $x_1 = -1$.
To describe a collections of trees that are separated but not
read-once, we consider (disjoint) compositions.
A \emph{disjoint composition} is a function
$F=f(f_1,\dots,f_m)$ where each $f_j$ acts on a disjoint set of
input variables, and the value of each of the input variables $x_j$
of $f$ is the value of $f_j$.
An example is given by Tribes (OR of disjoint ANDs).
It should be clear that a representation of a function
as a disjoint composition $F=f(f_1,\dots,f_m)$
together with a DDT for each factor
function $f,f_1,\dots,f_m$ induces a DDT for the composition;
one just needs to replace each node of the tree computing $f$
by a corresponding tree computing a function $f_j$.
It is not too hard to check that if each of the original trees
is separated, then also the tree calculating $F(f_1,\dots,f_m)$
is separated. In particular, recursive disjoint compositions of
read-once trees are separated. On the other hand,
it is easy to see that the simplest nontrivial Tribes function
$(x_1\wedge x_2)\vee(x_3\wedge x_4)$ cannot be represented
by a read-once tree.

\medskip

Finally, we discuss the necessity of the factor $k$ in
Corollary~\ref{cor:real}.  Indeed, as far as we know, it may be possible to
replace the factor $k$ by an absolute constant.
However we \emph{can} show that the factor $k$ cannot be replaced
by $1$. The $\bits^3 \to (\R, \rho_2)$ example shown in Figure~1
demonstrates that a constant
slightly greater than 1 is necessary. Except for optimizing the
leaf labels in this particular tree, this is the worst example we
know.

\begin{figure*}[t]
\begin{center}
\setlength{\unitlength}{1in}
\begin{picture}(4.0,1)(1,-.15)
\put(1,0){\epsfig{file=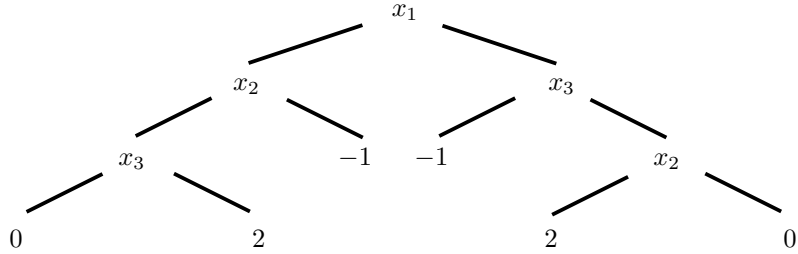,width=4.0in}}

\put(2.93,1.06){$x_1$} \put(2.1,.67){$x_2$} \put(3.75,.67){$x_3$}
\put(1.5,.28){$x_3$} \put(2.65,.28){$-1$} \put(3.05,.28){$-1$}
\put(4.3,.28){$x_2$} \put(.93,-.15){$0$} \put(2.2,-.15){$2$}
\put(3.73,-.15){$2$} \put(4.98,-.15){$0$}
\end{picture}
\caption{Left edges correspond to input variables with value $-1$,
right edges to value $1$. The
function $f:\bits^3 \rightarrow \R$ computed by this DDT has
$\Var[f]={\frac 3 2}$, but $(\delta_1(T),\delta_2(T),\delta_3(T))
= (1, {\frac 3 4}, {\frac 3 4})$ and
$(\Inf^{\rho_2}_1(f),\Inf^{\rho_2}_2(f),\Inf^{\rho_2}_3(f))=
({\frac 1 8}, {\frac 7 8}, {\frac 7 8})$, where
$\rho_2(x,y)=(x-y)^2/2$, so $\sum_{i=1}^3
\delta_i(T)\;\Inf^{\rho_2}_i(f) = {\frac {23} {16}} < \frac32$.}
\end{center}
\end{figure*}

\section{Questions for Future Work}
\label{sec:conclusions}

\begin{itemize}
 \item Is it possible to explain the ``coincidence'' that our
 near-tight lower bound on $\Deltap(f)$ for monotone transitive
 functions gives a lower bound for graph properties --- about $v^{4/3}$ --- that
 essentially matches the lower bound barrier that has stood since
 Hajnal '91~\cite{Hajnal:91}?  Perhaps either the
 Hajnal or the Chakrabarti-Khot~\cite{ChakrabartiKhot:01}
arguments can be reframed in terms of
 merely transitive functions (if true of Chakrabarti-Khot, this
 would be quite interesting); or, perhaps graph-theoretic
 arguments can augment our elementary probabilistic reasoning to
 produce a better lower bound.  
 \item Can our inequality in the real-valued, $\rho_2$ case ---
 Corollary~\ref{cor:real} --- be sharpened?  If the factor $k$ could
 be replaced by a universal constant, this would be a very strong
 variant of the Efron-Stein inequality.
 \item What other applications might our main inequality have?  We
 suggest there might be applications in
 computational learning theory or in the theory of random graphs.
\end{itemize}

\section{Acknowledgments}
We would like to thank Andris Ambainis, Laci Lovasz, and Avi
Wigderson for helpful discussions.

\bibliographystyle{latex8}
\bibliography{allrefs}

\end{document}